# Propagation Performance of Terahertz Channels in Lunar Dust

Peian Li,[1,2] Jiabiao Zhao,[1,2] Mingxia Zhang,[1,2] Yuheng Song,[1,2] Wenbo Liu,[1,2] Lingfeng Tian,[1,2] and Jianjun Ma[1,2,3*]

[1]*School of Integrated Circuits and Electronics, Beijing Institute of Technology, Beijing 100081, China*
[2]*Beijing Key Laboratory of Millimeter and Terahertz Wave Technology, Beijing 100081, China*
[3]*Tangshan Research Institute, BIT, Tangshan, Hebei 063099, China*

**Abstract:** The growing lunar exploration programs require robust communication systems for dust-laden environments, necessitating comprehensive understanding of channel propagation characteristics. We present an analysis of terahertz channel propagation through lunar dust environments, critical for reliable communication and sensing infrastructure. We develop an extended Mie scattering model incorporating unique properties of lunar dust particles from Apollo samples (10084, 14003, 70051), including irregular morphology, dielectric characteristics, and charge-dependent behavior. Through theoretical analysis and experimental verification, we examine power and bit error rate performance across varying dust conditions, revealing distinct relationships between particle characteristics and channel performance.

## 1. Introduction

Space exploration stands at a pivotal juncture, with lunar colonization emerging as humanity's next frontier in establishing a sustainable presence beyond Earth [1, 2]. The Moon's strategic position has catalyzed unprecedented international collaboration, exemplified by NASA's Artemis program's focus on scalable logistics and life-support systems [3], China's Chang'E program's advances in robotic exploration [4, 5], and the European Space Agency and JAXA's developments in lunar habitation technologies [6]. The strategic importance of lunar base establishment in China was prominently highlighted at the 759th Xiangshan Science Conference (China), which specifically emphasized the need for sensing and communication platforms in lunar exploration. This conference underscored China's commitment to developing comprehensive space environment platforms that combine environmental monitoring with high-bandwidth communication capabilities [7]. However, establishing sustainable lunar bases presents formidable technical challenges (demanding solutions for environmental monitoring and communication systems) [8, 9], particularly regarding lunar dust particles formed through micrometeorite impacts and space weathering. These particles exhibit distinctive characteristics including high surface area to volume ratios, jagged morphologies, and substantial electrostatic charging due to solar wind and UV radiation exposure [10-12]. While infrared systems have proven valuable for thermal mapping and compositional analysis, as demonstrated by the Lunar Reconnaissance Orbiter's Diviner radiometer [13, 14], their effectiveness for integrated sensing and communication is considerably diminished when encountering suspended dust particles (dust-laden environments) [15]. This necessitates the exploration of alternative frequencies capable of supporting both sensing and communication functions for reliable lunar infrastructure.

Terahertz (THz) technology, operating in the 0.1-10 THz frequency range, emerges as a promising solution for integrated lunar communication and sensing challenges [15, 16]. THz waves offer unique advantages for dual-use applications: their wavelengths are sufficiently short to enable both high-bandwidth communication (potentially exceeding 100 Gbps) and high-resolution sensing, yet long enough to achieve reasonable penetration through dust clouds. This inherent dual capability makes THz systems particularly suitable for establishing reliable communication infrastructure while simultaneously enabling real-time dust monitoring and surface characterization through the same hardware platform [17]. Fig. 1 illustrates a comprehensive lunar base THz infrastructure that integrates communication and sensing capabilities critical for sustained lunar operations. The multi-layered architecture encompasses satellite relay systems, surface-based communication towers, and local area networks operating in the THz frequency range, enabling high-bandwidth connectivity between the lunar base habitat, processing stations, and remote units.

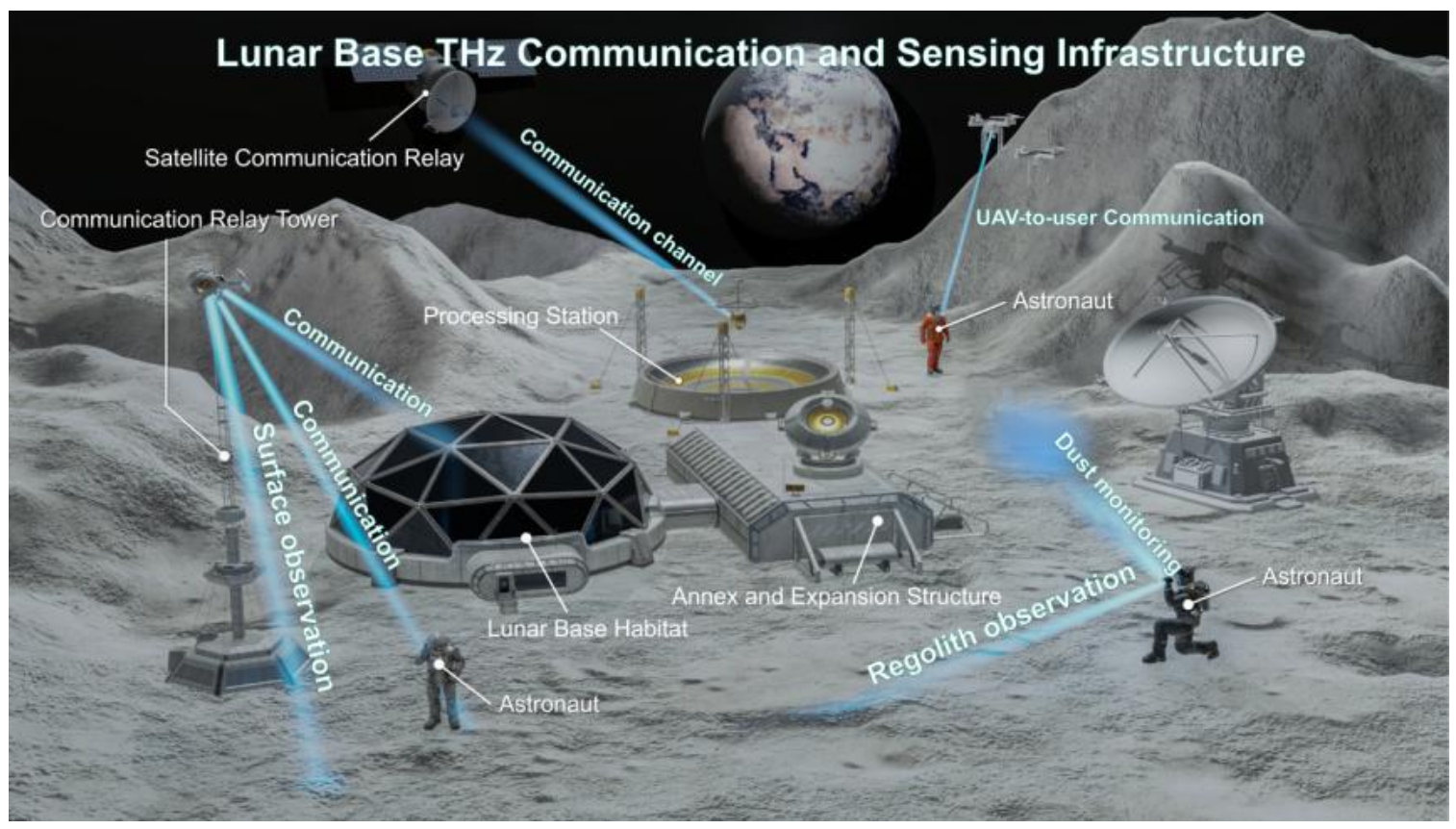

Fig. 1. Lunar base THz communication and sensing infrastructure. (The infrastructure incorporates dual-use THz transceivers that simultaneously support high-speed data transfer and environmental monitoring, including real-time dust particle characterization and regolith observation. This integrated approach, featuring redundant communication pathways and distributed sensing capabilities, ensures robust system performance while enabling scalable lunar base development.)

Recent research has significantly advanced our understanding of integrated THz applications in dust-laden environments. Studies indicate that while dust scattering affects THz channels, they demonstrate superior resilience for both sensing and communication channels compared to traditional RF channels, particularly at higher frequencies [18, 19]. Experimental investigations of THz interactions with lunar regolith analogs have revealed promising results in simultaneous surface reflection measurement and communication capabilities, critical for integrated remote sensing applications [20]. Advanced applications, such as THz computed tomography (THz-CT), demonstrate the potential for combining structural monitoring with communication links, enabling comprehensive infrastructure management essential for long-term habitation [21]. Furthermore, theoretical modeling of THz propagation through charged dust environments suggests viable performance for concurrent high-data-rate communication and environmental monitoring under lunar conditions [22, 23].

Despite these advances, significant gaps remain in our understanding of THz channel behavior in lunar environments, particularly regarding power profiles and bit error rate (BER) performance under varying dust conditions. Building upon our previous experimental and theoretical investigations [15, 16], this work aims to comprehensively characterize THz channel performance through lunar dust, with specific emphasis on quantifying signal degradation mechanisms and establishing performance boundaries for reliable communication links. Our analysis focuses on developing practical guidelines for the design and implementation of lunar dust sensing systems and wireless communication networks, contributing to the broader goal of establishing sustainable lunar infrastructure.

## 2. Channel model

### *2.1 Particle size distribution*

The modeling of THz channel propagation through lunar dust requires accurate characterization of particle size distributions. While many investigations employ theoretical models assuming spherical particles[24, 25], lunar dust particles exhibit highly irregular and angular morphologies due to space weathering and micrometeorite impacts. These characteristics necessitate more sophisticated approaches such as the Discrete Dipole Approximation, which can model arbitrary particle geometries by representing them as arrays of polarizable points, or the T-matrix method that provides exact solutions for rotationally symmetric particles. The effective medium approach offers another alternative by treating the dust-laden environment as a composite material with effective electromagnetic properties. The particle size distribution of dust particles can be expressed through the probability density function as

$$N(D) = \frac{N_0}{\sqrt{2\pi}\sigma D}\exp\left[-\frac{(\ln D - m)^2}{2\sigma^2}\right] \tag{1}$$

where $N_0$ represents the dust particle number density, $D$ is the particle diameter, $\sigma$ denotes the standard deviation of the log-transformed diameter, and $m$ is the mean of the log-transformed diameter. While this log-normal distribution has been validated for terrestrial environments like the Taklimakan desert [26], its direct application to lunar conditions requires careful consideration. The fundamental differences in particle formation mechanisms, environmental forces, and charging conditions between terrestrial and lunar environments make such extrapolations problematic [27, 28].

Analysis of Apollo mission samples has provided crucial insights into lunar dust characteristics. Measurements revealed distinct parameter values for Apollo 11 sample 10084 ($\sigma$ = 1.507, $m$ = -0.034), Apollo 14 sample 14003 ($\sigma$ = 0.789, $m$ = -0.025), and Apollo 17 sample 70051 ($\sigma$ = 1.200, $m$ = 0.125) [27, 28]. Importantly, we have not determined $N_0$ as these measurements are based on lunar regolith samples, which can only establish the probability density function but cannot quantify the actual number of particles in suspended lunar dust. The dust particle distributions are shown in Fig. 2(a) with the parameters for lunar dust near the lunar surface (regolith) [29, 30]. The diversity in sample characteristics reflects varying geological and environmental conditions across lunar regions, significantly impacting dust composition, particle sizes, and their interactions with THz channels. This variability, analogous to the effects observed in terrestrial precipitation (rain and snow) [25, 31], reinforces the value of Mie/Rayleigh scattering theory as a more general theoretical model for predicting channel power and bit error rate (BER) performance compared to empirical approaches.

### *2.2 Dielectric property*

The propagation characteristics of THz channels through the lunar environment are fundamentally governed by the dielectric properties of lunar dust particles. The complex relative permittivity $\varepsilon_r = \varepsilon_r' - j\varepsilon_r''$ serves as the primary descriptor of this interaction, where the real component $\varepsilon_r'$ represents the it's energy storage capacity, and the imaginary component $\varepsilon_r''$ quantifies energy dissipation. In lunar dust, these dielectric properties are substantially influenced by its mineral composition, particularly the concentrations of titanium dioxide ($TiO_2$) and iron oxide (FeO). Analysis of lunar regolith samples has established a regression model demonstrating that the loss tangent ( $\tan\delta = \varepsilon_r^{"}/\varepsilon_r^{'} + \sigma_0/\left(\omega\varepsilon_0\varepsilon_r^{'}\right)$ ) exhibits direct proportionality to both the combined abundance of $TiO_2$ and FeO (denoted as $S$ = %$TiO_2$ + %FeO) and the bulk density $\rho_k$ [32, 33]. This relationship is expressed as

$$\tan\delta = 10^{0.38S+0.312\rho_k-3.260} \tag{2}$$

where $\rho_k$ is dust bulk density measured in the range of 0.8 and 3.2 g/cm$^3$ [34, 35]. This relationship is particularly significant for THz channel propagation, as these mineral components can substantially affect signal attenuation. Studies utilizing data from the Moon Mineralogy Mapper (M3) and Apollo samples consistently demonstrate that $TiO_2$, in particular, exerts a dominant influence on the loss tangent [12]. The abundance parameter $S$ typically varies between 4-32% [34], with 14% selected as a representative value for this work. The real part of the dielectric constant demonstrates a strong dependence on bulk density, following the relationship

$$\varepsilon_r^{'} = 1.919^{\rho_k} \tag{3}$$

which shows remarkably little dependence on chemical or mineralogical composition, making density as the primary determining factor for the real component of the dielectric constant.

Lunar dust particles undergo significant modifications to their dielectric properties through multiple charging mechanisms in the lunar environment. When exposed to solar ultraviolet and X-ray radiation, particularly intense on the Moon's dayside, these particles experience the photoelectric effect, which causes electron ejection and results in positive charging [36]. The charging process is further enhanced by high-energy solar wind electrons and cosmic rays impacting the lunar surface, triggering secondary electron emission [37]. These combined charging mechanisms fundamentally alter both the particles' permittivity and electrical conductivity, with measurements from Apollo 15 samples (15301,38) revealing a distinct temperature-dependent conductivity pattern [38], as

$$\sigma_0 = 6\times10^{-18}e^{0.0237T} \tag{4}$$

where $T$ represents temperature in Kelvin, ranging from 50K to 400K [39]. This temperature-dependent conductivity reflects thermally activated hopping conduction between localized states in amorphous silicate matrices, characteristic of radiation-damaged lunar soil particles. The diurnal temperature fluctuations induce order-of-magnitude variations in conductivity, which significantly affects charge dissipation rates. The charging effect modifies the dielectric constant of lunar dust [34, 40, 41] according to

$$\varepsilon_r^{'} = 1.919^{\rho_k}\left(1+\alpha Q\rho_k\right) \tag{5}$$

where $\alpha$ represents the charge-polarizability coefficient and $Q$ denotes the total charge on a single dust particle. The conductivity modification, based on the Drude model, becomes

$$\sigma_0 = 6\times10^{-18}e^{0.0237T}\left(1+\frac{\tau_D}{1+\omega^2\tau_d^2}\cdot\frac{nq^2}{m_e}\right) \tag{6}$$

Here, $\tau_D$ represents the reference relaxation time, $\tau_d$ is the adjusted relaxation time accounting for irregular particle shapes, $q$ denotes the electron charge, and $m_e$ is the electron mass. The charge carrier density $n$ is determined by both the lunar dust particle density ($N_0$) and charge number ($N_e$) per particle. While charge numbers can reach up to $10^5$ per particle, a conservative value of $N_e$ = 1000 per particle is adopted in this work [42].

The impact of charging on dielectric properties exhibits threshold behavior, as shown in Fig. 2(b). For charge numbers below $10^3$ per particle, the real part of the dielectric constant remains relatively stable, with values consistent with measurements at the Chang'E-5 (2.52) [43]. However, when $N_e$ exceeds $10^3$ per particle, $\varepsilon_r^{'}$ increases significantly to measurements at the Chang'E-3 (2.89) landing sites [43], due to enhanced polarization effects. This threshold behavior suggests that only sufficient charge accumulation can meaningfully influence the particles' polarization response to electric fields [44].

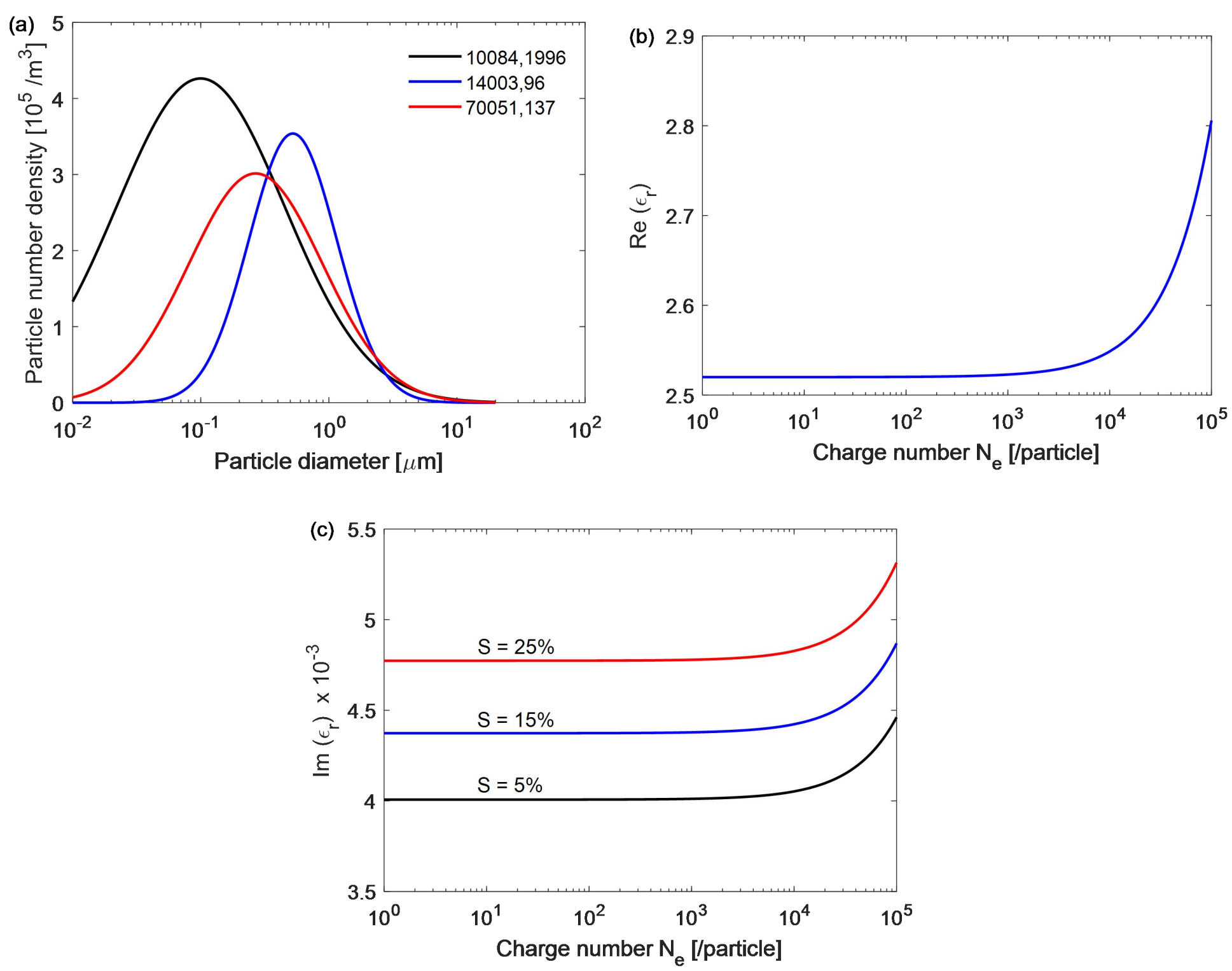

Fig. 2. (a) Lunar dust particle size distribution for different samples. (b) Variation of real relative permittivity of the charged lunar dust particles with respect to charge number. (c) Variation of imaginary relative permittivity of the charged lunar dust particles with respect to charge number and . (Parameter setting as in Table 1)

The imaginary part of the dielectric constant ( $\varepsilon_r^{''}$ ) exhibits interesting behavior under varying conditions, as shown in Fig. 2(c). For concentrations of $S$ = 25% ($TiO_2$ + FeO), $\varepsilon_r^{''}$ remains relatively constant of $\varepsilon_r^{''} = 4.73\times10^{-3}$ when the charge number stays below $10^3$ per particle [45]. The low conductivity and minimal dielectric losses of lunar dust particles contribute to their exceptional charge retention capabilities. This characteristic enables dust particles to maintain their electrical charge for extended periods, fundamentally affecting their interaction with THz channels over time. Notably, this component ( $\varepsilon_r^{''}$ ) shows negligible dependence on both temperature and operating frequency, suggesting that while the concentration of $TiO_2$ + FeO influences the loss factor, its effect on power loss remains modest [46]. The diurnal cycle introduces significant variations in dust behavior through its impact on photoelectric charging and temperature-dependent conductivity. During lunar sunrise and sunset, the substantial photo-induced changes in electrical conductivity can charge dust particles sufficiently to induce levitation and movement. Individual dust particles can acquire substantial surface charges, ranging from $10^4$ to $10^5$ elementary charges [47], through combined photoelectric emission and solar wind interactions. This charging particularly affects particles and modifies their effective scattering cross-sections and, consequently, their interaction with THz channel.

**Table 1 Parameter setting for channel modeling.**

| Parameter | Value |
|---|---|
| Operating frequency | 1.5 [THz] |
| Temperature | 220 [K] |
| Bulk density | 1.418 [g/cm3] |
| Dust particle density | 5.0×105 [/m3] |
| Abundance S of $TiO_2$ + FeO | 14% |
| Charge number | 1000 [/particle] |
| Correlation factor | 0.6 |
| Transmitting power | 25 dBm |
| T/R antenna gain | 45 dBi |
| Noise level | -60 dBm |
| Channel distance | 1 km |

### *2.3 Propagation theory*

The propagation of THz channels through lunar dust presents unique theoretical challenges that require careful consideration of both particle characteristics and channel-particle interactions. The Rayleigh approximation is valid only when the particle size is much smaller than the wavelength of the incident wave, such as less than a twentieth of the wavelength [48]. Additionally, it does not incorporate the effects of charge excitation on dust particles [49]. So in this work we used the Mie scattering theory for channel modeling and our previous experimental work with volcanic dust (average radius 4.3 μm) has demonstrated the efficacy of Mie theory for modeling 625 GHz channel behavior [15, 16]. However, applying this framework to the lunar environment necessitates important modifications to account for the distinctive properties of lunar dust particles.

The classical Mie scattering theory, when applied to dust-laden environments, relies on two fundamental assumptions, as negligible multiple scattering effects and independent behavior of individual scatterers [50]. Under these conditions, the attenuation experienced by a THz channel propagating through a dust-filled medium can be expressed as the formula

$$\alpha_{dust} = 4.343 \cdot 10^3 \int_0^\infty \xi_e\left(n_{eff}, \chi\right) \cdot N\left(r\right) \cdot \pi r^2 dr \tag{7}$$

where $r$ represents the particle radius, $N(r)$ denotes the particle number density as Eq. (1), and $\xi_e\,(n_d, \chi)$ is the extinction efficiency. This efficiency parameter can be expanded as

$$\xi_e\left(n_d, \chi\right) = \frac{2}{\chi^2} \sum_{l=1}^{\infty} \left(2l+1\right) \mathrm{Re}\left(a_l + b_l\right) \tag{8}$$

Here, $\chi = 2\pi r/\lambda_0$ represents the normalized circumference, with $\lambda_0$ being the free-space wavelength, and $a_l$ and $b_l$ are scattering coefficients derived from the infinite series expansion of the scattered electromagnetic field [16]. These coefficients depend on the relative refractive index $n_d$ of the lunar dust particles. Experimental validation of this theoretical framework has shown good agreement with measurements conducted in volcanic dust environments, particularly for particles following a Gaussian distribution with an average radius of 4.3 µm [15]. Fig.3 illustrates the comparative analysis between theoretical predictions and measured data for both THz and IR channels, demonstrating the model's accuracy for spherical particles with refractive indices of 1.54 at 625 GHz and 1.5 in the IR range.

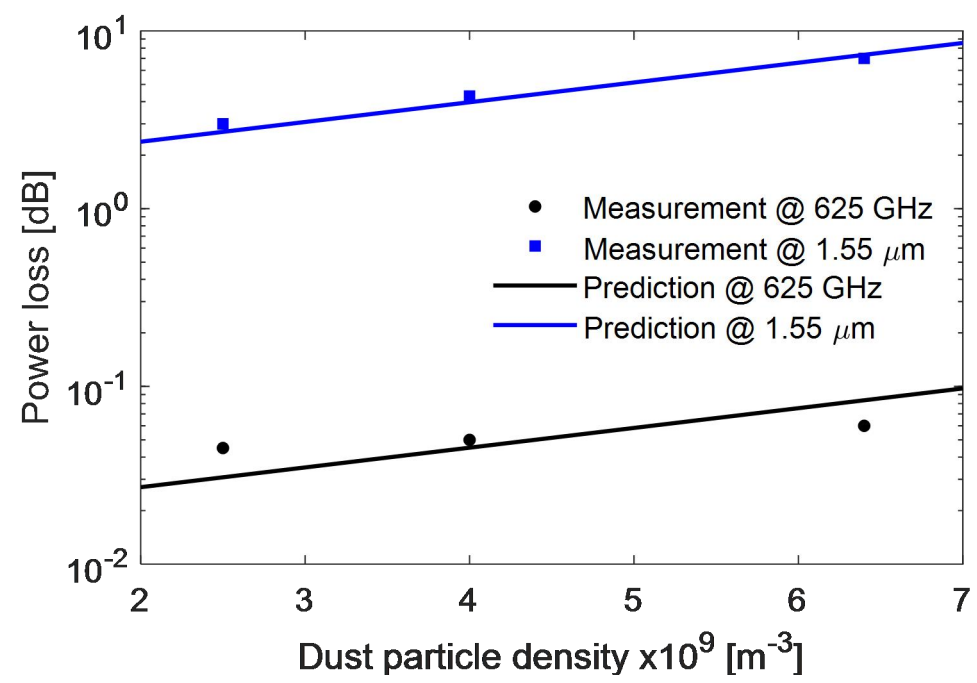

Fig. 3. Power loss suffered by THz and IR channel in volcanic dust ( composed primarily of montmorillonite and beidellite) in measurement and theoretical prediction. (refractive index is 1.54 at 625 GHz and 1.5 in the IR range).

However, the application of classical Mie scattering theory to lunar dust requires significant modifications to account for the distinctly non-spherical nature of lunar dust particles. Unlike volcanic dust particles that can be reasonably approximated as spheres, lunar dust particles exhibit extreme morphological irregularity, characterized by sharp edges, complex conglomerates formed through high-temperature sintering, and occasional nearly-spherical droplets [37]. This morphological complexity stems from the explosive origin of lunar dust and subsequent space weathering processes. To adapt the Mie scattering framework for these irregular particles, the extinction efficiency $\xi_e$ ($n_d$, $\chi$) requires modification. The adapted form introduces a shape correction factor $k_s$:

$$\xi_e\left(n_d, \chi\right) = \xi_e\left(n_d, \chi\right) \cdot k_s \tag{9}$$

where $k_s$ varies within the range of 0.5-1 [51], based on both theoretical considerations and experimental observations. The lower bound ($k_s$=0.5) represents cases where particle irregularity significantly reduces scattering efficiency compared to equivalent-volume spheres, while the upper bound ($k_s$=10) corresponds to situations where particle shape effects are minimal.

## 3. Power performance

Unlike terrestrial environments, the Moon's lack of substantial atmosphere or ionosphere creates distinctive conditions for THz wave propagation [52]. The absence of atmospheric interference that typically causes significant attenuation and refraction on Earth [53], suggests the potential for THz channels to traverse considerably longer distances in lunar environments. This characteristic shifts the primary focus of power performance analysis to the interaction between THz channels and lunar dust particles, as dust becomes the dominant factor affecting channel propagation.

Theoretical analysis conducted on three distinct lunar dust samples (see Fig. 4(a)) reveals characteristic power loss patterns in the THz frequency range, with scattering emerging as the dominant loss mechanism. Similar phenomenon has been observed in reference [54].The comparative analysis identifies the 2.4 THz frequency band as particularly sensitive to lunar dust interactions. This heightened sensitivity at 2.4 THz presents both challenges and opportunities for lunar communications and sensing applications- while it results in increased power loss that must be compensated for in communication system design, it also enables more precise characterization of dust environments through enhanced interaction with particles.

Importantly, Fig. 4(b) demonstrates that the backscattering loss is substantially lower than the total (scattering) loss, indicating that dust particles predominantly scatter THz waves in the forward direction. This directional scattering behavior suggests that integrated sensing and communication systems should utilize forward-propagating channels between transmitter-receiver pairs, rather than relying on backscattered signals. Such a configuration would not only provide enhanced sensitivity for dust characterization but also achieve superior energy efficiency by leveraging the natural forward-scattering properties of lunar dust. The power loss measurements across different frequencies and dust samples provide crucial insights for optimizing these systems. Notably, the variation in power loss with respect to dust particle density spans several orders of magnitude as particle concentration increases from $10^4$ to $10^7$ particles per unit volume. This extensive dynamic range presents significant challenges for communication system design, necessitating sophisticated power detection capabilities, advanced linearization techniques, and robust signal processing methods to maintain reliable communication links in varying dust conditions.

The charging of lunar dust particles introduces complex modifications to their interaction with THz waves, primarily through alterations in their polarization response to electromagnetic fields. As previously established in Fig. 2(b), charged dust particles exhibit enhanced polarization capabilities when subjected to external electric fields. Our analysis in Fig 4(c) reveals a notable inverse relationship between particle charge levels and power loss characteristics, particularly as charge density approaches $N_e = 10^5$ particles per particle. As particle charge numbers increase, we observe a systematic shift in frequency sensitivity toward lower frequencies (detailed in Supplementary Information Fig. S1). This phenomenon likely stems from charge-induced modifications to the dust particles' dielectric properties, fundamentally altering their THz wave interactions [44, 55].

The morphological characteristics of lunar dust particles, quantified through the correlation factor $k_s$, significantly influence power loss patterns. Our analysis demonstrates a clear inverse relationship between $k_s$ values and power loss magnitude, as shown in Fig. 4(c). Irregular particles (lower $k_s$ values) exhibit highly orientation-dependent scattering cross-sections. This variability introduces additional complexity in scattering behavior due to surface roughness effects, particle orientation dynamics and enhanced diffraction effects [56]. Besides, the increasing irregularity of particles leads to greater discrepancies between assumed and actual particle size distributions, which could also contribute to the observed reduction in power loss at lower $k_s$ values.

Environmental factors, specifically temperature ($T$) and abundance ($S$) of $TiO_2$ and FeO, demonstrate surprisingly minimal impact on power loss characteristics (detailed in Supplementary Information Fig. S2). This relative insensitivity to compositional variations (within the 0-30% abundance range) and temperature fluctuations suggests that these parameters may be considered secondary factors in THz channel modeling [46, 57]. This finding significantly simplifies certain aspects of dust modeling and system design, allowing focus on more critical parameters such as particle charge and morphology.

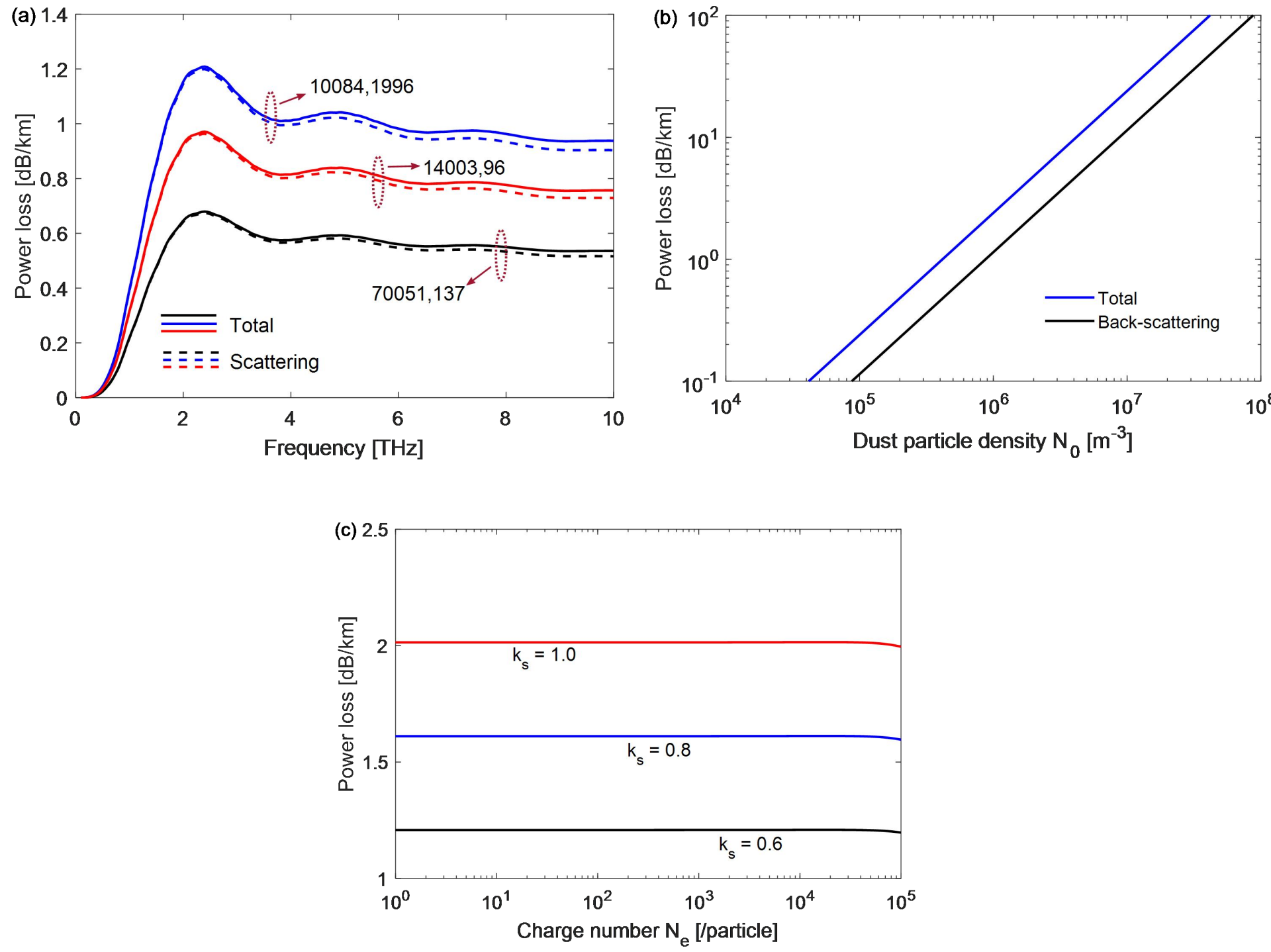

Fig. 4. (a) Variation of power loss with respect to THz frequency through different lunar dust samples. (b) Variation of total and back-scattering power loss with respect to dust particle density. (c) Variation of power loss with respect to charge number under different correction factors (ks). Parameter setting as Table 1.

## 4. BER performance

The reliability of THz communication systems in lunar environments can be comprehensively evaluated through BER analysis, which provides direct insight into the quality of received signals by quantifying transmission errors caused by noise, interference, and channel distortion. While power loss analysis characterizes overall signal attenuation, BER assessment is crucial for optimizing system parameters and error correction mechanisms to maintain reliable communication channels while preserving sensing capabilities. The BER performance analysis in lunar dust environments requires consideration of both the unique characteristics of THz channel propagation and the specific properties of lunar dust particles.

In our analysis, we employ quadrature phase-shift keying (QPSK) modulation due to its superior spectral efficiency and robust performance in noise-limited channels [31]. Our simulation framework implements a comprehensive model of THz channel propagation through lunar dust using MATLAB R2024a. The simulation incorporates the established theoretical foundations while accounting for the unique characteristics of lunar dust particles. Due to the Moon's extremely thin atmosphere, approximately $10^{15}$ times less dense than Earth's atmosphere in terms of molecules per cubic centimeter [35], we neglect atmospheric absorption effects in our model. The free space path loss is calculated using standard propagation models (the 'fspl' function in MATLAB), with particular attention paid to the frequency-dependent characteristics at 2.4 THz, our chosen operating frequency due to its enhanced sensitivity to lunar dust interactions.

Initial simulations utilizing MATLAB R2024a's 'bertool' functionality were conducted with the system parameters specified in Table 1, including a carrier frequency of 2.4 THz, transmit power of 25 dBm, antenna gains of 45 dBi and noise level of -60 dBm. Our comparative analysis of Rician and AWGN channel models reveals distinct behavioral patterns in the presence of lunar dust. For Rician channels with K-factor = 10, we observe significant BER degradation compared to the AWGN case, as illustrated in Fig. 5(a). The degradation manifests primarily through increased error rates at higher SNR values, indicating the dominant effect of multipath components. However, when the K-factor increases to 100, the BER performance converges remarkably close to that of an AWGN channel, suggesting minimal multipath interference. This behavior demonstrates interesting parallels with our previous findings in terrestrial precipitation studies [19], where K-factors exceeding 40 dB were observed in rain and snow conditions. Given the similarities in scattering mechanisms between suspended lunar dust particles and terrestrial precipitation, coupled with the Moon's lack of atmospheric turbulence, we anticipate K-factors in lunar dust environments to fall within a similar range. The expected high K-factors in lunar environments arise from the predominance of line-of-sight components and relatively ordered dust particle distributions, yielding performance characteristics nearly identical to AWGN channels. This observation provides relative strong justification for our subsequent focus on AWGN channel modeling in lunar dust environments.

Furthermore, our analysis reveals subtle but significant variations in BER performance across different lunar dust samples. Detailed examination of Apollo mission samples, particularly samples 10084, 14003, and 70051, demonstrates distinct BER characteristics that correlate with their unique particle size distributions and morphological properties. While these variations may appear modest in conventional BER measurements, they become particularly pronounced when utilizing high-precision BER testing equipment capable of resolving error rates below $10^{-10}$. This enhanced sensitivity to dust composition suggests promising applications for dust characterization through BER pattern analysis, potentially enabling real-time monitoring of lunar dust conditions through communication system performance metrics.

Particle density emerges as a critical factor in determining BER performance, as demonstrated in Fig. 5(b). Our results show dramatic BER variations as dust particle density increases from $10^4$ to $10^7$ particles per unit volume. This relationship is further complicated by particle morphology, quantified through the correction factor $k_s$. The strong correlation between BER patterns and particle characteristics suggests that BER analysis could serve as a method for characterizing average lunar dust particle shape, providing valuable environmental monitoring capabilities alongside communication functions. A particularly concerning observation is the system's approach to an error floor at high dust densities. Under these conditions, the BER becomes increasingly insensitive to SNR improvements, indicating that traditional methods of enhancing performance through increased transmission power or more robust error correction coding may prove ineffective [58]. For practical implementations, we recommend maintaining link margins of at least 20 dB above the minimum required SNR for the desired BER performance under clear conditions. This additional margin provides headroom for dust-induced degradation while maintaining adequate sensing sensitivity [59]. However, system designers should note that this margin becomes less effective at high dust densities due to the error floor effect, emphasizing the importance of implementing the previously mentioned adaptive techniques. Future lunar communication infrastructure should also incorporate redundant paths and backup systems to ensure continuous connectivity during severe dust events [60, 61], particularly for critical operations and environmental monitoring functions.

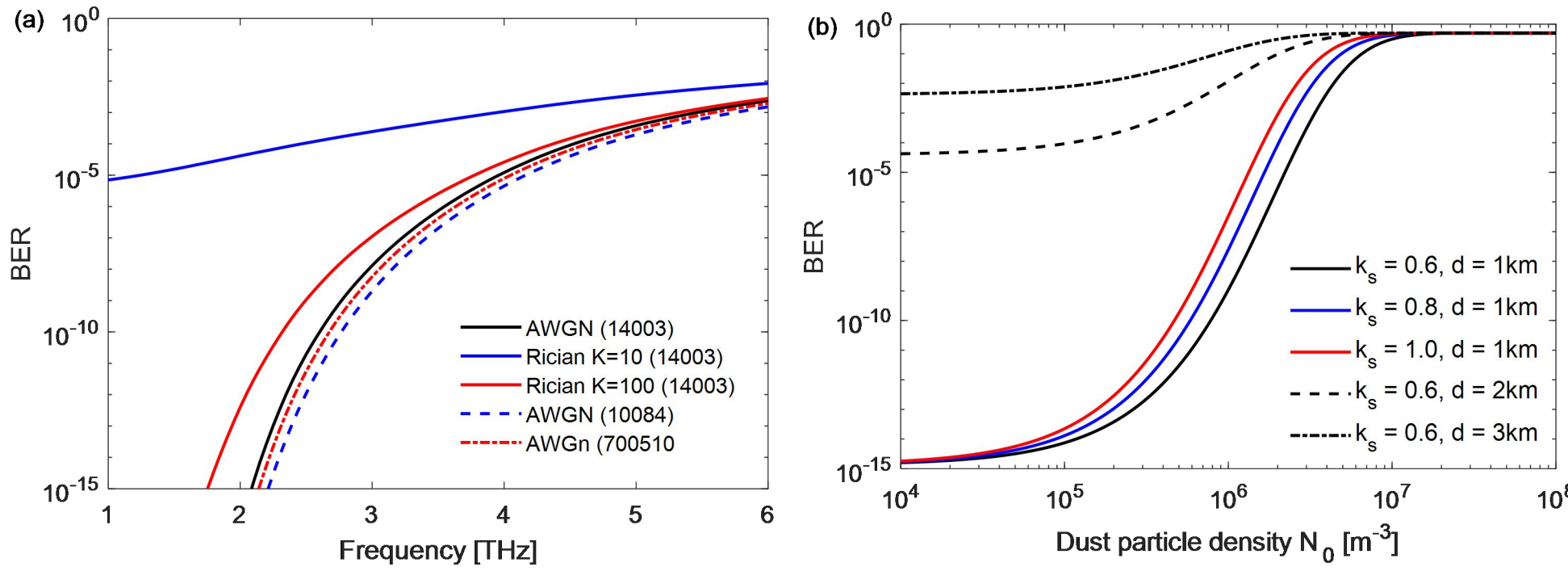

Fig. 5. Variation of BER with respect to (a) dust particle density for three kinds of lunar samples, and (b) dust particle density for different correlation factors and distances

## 5. Conclusion

The establishment of sustainable lunar bases and reliable communication infrastructure represents a critical challenge in current space exploration initiatives, with lunar dust emerging as a significant impediment to conventional communication technologies. In this work, we present a comprehensive investigation of THz channel propagation through lunar dust environments, contributing fundamental insights essential for developing robust lunar communication and sensing infrastructure. Our analysis revealed that the 2.4 THz frequency band demonstrates particular sensitivity to lunar dust interactions with strong scattering characteristics. We also observed an inverse relationship exists between particle charge levels and power loss as charge density approaches $10^5$ particles per particle. Besides, irregular particle morphology (quantified by the correlation factor) significantly influences power loss through orientation-dependent scattering. Notably, environmental factors such as temperature and $TiO_2$/FeO abundance showed minimal impact on power loss within typical ranges. Our BER analysis demonstrated that Rician channels with high K-factor converge to AWGN channel behavior, while while three distinct Apollo mission dust samples (Apollo 11 sample 10084, Apollo 14 sample 14003, and Apollo 17 sample 70051) exhibited unique BER characteristics correlating with their specific particle properties and distributions.

These findings not only advance our understanding of THz channel performance in lunar environments but also provide concrete guidelines for the development of reliable communication systems capable of operating under challenging dust conditions. In this work, we do not take efforts on channel's sensing efficiency, which should be our future aspects. Additionally, future investigations should explore adaptive techniques for optimizing channel performance under varying dust conditions, dynamic frequency selection mechanisms for maintaining reliable links during dust storms, and enhanced modeling approaches that incorporate the complex interactions between multiple dust particles.

**Acknowledgment.** Jianjun Ma acknowledges support from the National Natural Science Foundation of China under Grant (62471033) and the Talent Support Program of Beijing Institute of Technology "Special Young Scholars" under Grant (3050011182153).

### References

1. C. Heinicke and B. A. C. H. J. S. Foing, "Human habitats: prospects for infrastructure supporting astronomy from the moon," Philosophical Transactions of the Royal Society A **379**(2188), 20190568 (2021).
2. M. Y. Marov and E. N. Slyuta, "Early steps toward the lunar base deployment: Some prospects," Acta Astronautica **181**, 28-39 (2021).

3. R. P. Mueller, "Lunar Base Construction Planning," in *18th Biennial International Conference on Engineering, Science, Construction, and Operations in Challenging Environments*, 2022), 858-870.
4. Y. Zheng, Z. Ouyang, C. Li, et al., "China's lunar exploration program: present and future," Planetary and Space Science **56**(7), 881-886 (2008).
5. C. Li, Chi Wang, Yong Wei, and Yangting Lin, "China's present and future lunar exploration program," Science **365**(6450), 238-239 (2019).
6. K. Matsumoto, N. Kamimori, Y. Takizawa, et al., "Japanese lunar exploration long-term plan," Acta Astronautica **59**(1-5), 68-76 (2006).
7. Q. Yang, J. Shen, R. Xu, et al., "Toward sustainable lunar base development: Comprehensive space environment platform," The Innovation Energy **2**(1), 100076 (2025).
8. A. Salmeri and M. Poliacek, "All for one and one for all: Recommendations for sustainable international lunar base utilization and exploration approaches," in *71st International Astronautical Congress - The Cyberspace Edition*, (International Astronautical Federation, 2020),
9. P. Zhang, W. Dai, R. Niu, et al., "Overview of the Lunar In Situ Resource Utilization Techniques for Future Lunar Missions," Space: Science & Technology **3**, 0037 (2023).
10. J. R. Gaier, *The Effects of Lunar Dust on EVA Systems During the Apollo Missions* (2005).
11. H. M. J. and J. Feighery, "Lunar Dust: Characterization and Mitigation," in *9th International Conference on the Exploration and Utilization of the Moon/International Lunar Exploration Working Group*, 2007),
12. P. B. Abel, M. D. Anderson, E. T. Blom, et al., *Lunar Dust Mitigation: A Guide and Reference: First Edition (2021)* (NASA Glenn Research Center Cleveland, Ohio, United States, 2023).
13. A. Bhattacharya, C. Li, N. O. Renno, et al., "Probing Dust and Water in Martian Atmosphere with Far-Infrared Frequency Spacecraft Occultation," Remote Sensing **15**(18), 4574 (2023).
14. W. Hu, Y. Bai, and X. Lv, *Space Terahertz Remote Sensing Technology* (CRC Press, Oxon, OX, 2024).
15. J. F. Federici, J. Ma, L. Moeller, et al., "Review of weather impact on outdoor terahertz wireless communication links," Nano Communication Networks **10**, 13-26 (2016).
16. K. Su, L. Moeller, R. B. Barat, et al., "Experimental comparison of terahertz and infrared data signal attenuation in dust clouds," J Opt Soc Am A **29**(11), 2360-2366 (2012).
17. S. Wang, "Terahertz Emission Modeling of Lunar Regolith," Remote Sensing **16**(21), 4037 (2024).
18. S. Noreen, F. Giannetti, V. Lottici , "Earth and Martian Sand and Dust Storms: A Comprehensive Review of Attenuation Modelling and Measurements," IEEE Access **12**, 878 - 922 (2023).
19. J. Ma, Y. Song, M. Zhang, et al., "Terahertz Channels in Atmospheric Conditions: Propagation Characteristics and Security Performance," Fundamental Research 5(2), 526-555 (2024).
20. Q. Rao, G. Xu, W. Mao, et al., "Detection of the Lunar Surface Soil Permittivity with Megahertz Electromagnetic Wave," Sensors **21**(7), 2466 (2021).
21. E. Z. Tucker, J. N. Zalameda, W. C. Wilson, et al., "Material Inspectability Assessment Related to Lunar Construction Technology Development," (2024).
22. L. T. Wedage, B. Butler, S. Balasubramaniam, et al., " Comparative analysis of terahertz propagation under dust storm conditions on mars and earth," IEEE Journal of Selected Topics in Signal Processing **17**(4), 745-760 (2023).
23. Z. Zhang and O. B. Akan, "Analysis of Terahertz Communication Under Dust Storm Conditions on Mars," IEEE Communications Letters **29**(2), 388-392 (2024).
24. P. Li, J. Wang, L. Zhao, et al., "Performance degradation of terahertz channels in emulated rain," Nano Communication Networks **35**, 100431 (2023).
25. P. Li, W. Liu, J. Liu, et al., "Measurement and Modeling on Terahertz Channels in Rain," Journal of Infrared, Millimeter, and Terahertz Waves **46**(1), 1-16 (2024).
26. M. Wang, H. Ming, Z. Ruan, et al., "Quantitative detection of mass concentration of sand-dust storms via wind-profiling radar and analysis of Z-M relationship," Theoretical and Applied Climatology **131**(3), 927-935 (2016).
27. J. Park, Y. Liu, K. D. Kihm, et al., "Characterization of Lunar Dust for Toxicological Studies. I: Particle Size Distribution," Journal of Aerospace Engineering **21**(4), 266-271 (2008).
28. D. S. McKay, B. L. Cooper, L. A. Taylor, et al., "Physicochemical properties of respirable-size lunar dust," Acta Astronautica **107**, 163-176 (2015).
29. D. A. Glenar, T. J. Stubbs, J. E. McCoy, et al., "A reanalysis of the Apollo light scattering observations, and implications for lunar exospheric dust," Planetary and Space Science **59**(14), 1695-1707 (2011).
30. D. R. Criswell, "Lunar dust motion," in *Proceedings of the Third Lunar Science Conference.*, (Pergamon Press, 1972) **3**, 2671-2680.
31. G. Liu, X. He, J. Zhao, et al., "Impact of snowfall on terahertz channel performance: measurement and modeling insights," IEEE Transactions on Terahertz Science and Technology **14**(5), 691-698 (2024).
32. Z. Meng and J. Ping, *Lunar Surface, Dielectric Permittivity* (Prairie View A&M University, Houston, USA, 2017), 1-3.
33. S. Wang, T. Yamada, K. S. Chen, et al., "Terahertz Scattering and Emission from the Lunar Surface," in *IEEE International Geoscience and Remote Sensing Symposium (IGARSS)*, IEEE International Symposium on Geoscience and Remote Sensing IGARSS (IEEE, 2022), 354-357.
34. D. W. Strangway, G. W. Pearce, G. R. Olhoeft, "Magnetic and dielectric properties of lunar samples," in *The Soviet-Am. Conf. on Cosmochem. of the Moon and Planets, pt. 1* 1977), 417-431.

35. G. H. Heiken, et al., *Lunar Sourcebook, a User's Guide to the Moon* (Cambridge University Press, Cambridge, UK, 1991).
36. J. E. Colwell, S. Batiste, M. Horányi, et al., "Lunar surface: Dust dynamics and regolith mechanics," Reviews of Geophysics **45**(2) (2007).
37. A. V. Zakharov, L. M. Zelenyi, S. I. Popel', "Lunar Dust: Properties and Potential Hazards," Solar System Research **54**(6), 455-476 (2020).
38. G. R. Olhoeft, A. L. Frisillo, D. W. Strangway, et al., "Temperature dependence of electrical conductivity and lunar temperatures," The Moon **9**(1), 79-87 (1974).
39. J.-P. Williams, D. A. Paige, B. T. Greenhagen, et al., "The global surface temperatures of the Moon as measured by the Diviner Lunar Radiometer Experiment," Icarus **283**, 300-325 (2017).
40. B. S. Ponto and J. C. Berg, "Charging of oxide nanoparticles in media of intermediate dielectric constant," Langmuir **35**(47), 15249-15256 (2019).
41. E. Bichoutskaia, A. L. Boatwright, A. Khachatourian, et al., "Electrostatic analysis of the interactions between charged particles of dielectric materials," The Journal of chemical physics **133**(2), 024105 (2010).
42. S. I. Popel, L. M. Zelenyi, A. Y. Dubinskii, "Lunar dust and dusty plasmas: Recent developments, advances, and unsolved problems," Planetary and Space Science **156**, 71-84 (2018).
43. Y. Li, B. Zhou, S. Shen, et al., "The Lunar Regolith Structure and Electromagnetic Properties of Chang'E-5 Landing Site," Remote Sensing **14**(18), 4539 (2022).
44. W. Liu, P. Li, D. Li, et al., "Transmission characteristics of millimeter and sub-terahertz channels through spatially ripple plasma sheath layers," IEEE Transactions on Plasma Science **52**(10), 5287-5295 (2024).
45. J. Feng, M. A. Siegler, M. N. White, "Dielectric properties and stratigraphy of regolith in the lunar South Pole-Aitken basin: Observations from the Lunar Penetrating Radar," Astronomy & Astrophysics **661**, A47 (2022).
46. L. Dihui, J. Jingshan, W. Ji, et al., "Experimental research and statistical analysis on the dielectric properties of lunar soil simulators," Chinese Science Bulletin **50**(10), 1034-1044 (2005).
47. M. Wang, W. Wei, Z. Ruan, et al., "Application of wind-profiling radar data to the analysis of dust weather in the Taklimakan Desert," Environmental Monitoring and Assessment **185**(6), 4819-4834 (2012).
48. A. Ishimaru, *Wave propagation and scattering in random media* (Academic Press, New York, 1978).
49. X. Li, L. Xingcai, Z. Xiaojing , "Attenuation of an electromagnetic wave by charged dust particles in a sandstorm," Appl. Opt. **49**(35), 6756-6761 (2010).
50. D. Deirmendjian, *Electromagnetic scattering on spherical polydispersions* (American Elsevier Publishing, New York, 1969).
51. M. Kocifaj and H. Horvath, "Inversion of extinction data for irregularly shaped particles," Atmospheric Environment **39**(8), 1481-1495 (2005).
52. J. Ma, Y. Song, M. Zhang, et al., "Terahertz Channels in Atmospheric Conditions: Propagation Characteristics and Security Performance," Fundamental Research **5**(2),526-555 (2024).
53. J. Federici and L. Moeller, "Review of terahertz and subterahertz wireless communications," J Appl Phys **107**(11), 111101 (2010).
54. Y. Amarasinghe, W. Zhang, R. Zhang, et al., "Scattering of Terahertz Waves by Snow," Journal of Infrared, Millimeter, and Terahertz Waves **41**(2), 215-224 (2019).
55. R. Su, Y. Zhao, J. Ma, et al., "Propagation characteristics of obliquely incident THz waves in inhomogeneous fully ionized dusty plasma using the scattering matrix method," Physics of Plasmas **31**(4), 043701 (2024).
56. T. Lurton, J. B.. Renard, D. Vignelles, et al., "Light scattering at small angles by atmospheric irregular particles: modelling and laboratory measurements," Atmospheric Measurement Techniques **7**(4), 931-939 (2014).
57. R. C. Anderson, M. Buehler, S. Seshadri, et al., "Dielectric constant measurements for characterizing lunar soils," Lunar and Planetary Science XXXVI, Part 1 (2005).
58. I. Korn, "Error floors in the satellite and land mobile channels," IEEE Transactions on Communications **39**(6), 833-837 (1991).
59. R. Narasimha, M. Lu, N. R. Shanbhag, et al., "BER-optimal analog-to-digital converters for communication links," IEEE transactions on signal processing **60**(7), 3683-3691 (2012).
60. K. Islam, W. Shen, and X. Wang, "Wireless sensor network reliability and security in factory automation: A survey," IEEE Transactions on Systems, Man, and Cybernetics, Part C (Applications and Reviews) **42**(6), 1243-1256 (2012).
61. J. C. Juarez, A. Dwivedi, A. R. Hammons, et al., "Free-space optical communications for next-generation military networks," IEEE Communications Magazine **44**(11), 46-51 (2006).